\documentclass[a4paper,fleqn,usenatbib]{mnras}

\usepackage[T1]{fontenc}
\usepackage{ae,aecompl}

\usepackage[english]{babel}
\usepackage{amsmath}	
\usepackage{amssymb}	

\usepackage{ulem}  

\usepackage{graphicx} 
\graphicspath{{figs/}}
\usepackage{rotating} 
\usepackage{caption}
\usepackage{booktabs}
\usepackage[flushleft]{threeparttable}

\newcommand\Mgas{M_{\rm gas}}
\newcommand\Mmax{M_{\rm max}}
\newcommand\Mstar{M_{*}}

\newcommand\Ms{{\cal M}_{\rm s}}
\newcommand\sfravg{\langle{\rm SFR}\rangle}

\newcommand\Mcl  {M_{\rm cl}}

\newcommand\Msun  {M_\odot}
\newcommand\Mtot  {M_{\rm tot}}

\newcommand\eff{\epsilon_{\rm ff}}
\newcommand\fden{f_{\rm d}}
\newcommand\fmol{f_{\rm mol}}

\newcommand\kms{\rm \, km ~ s^{-1}}

\newcommand\nmol{n_{\rm mol}}
\newcommand\nsf{n_{\rm SF}}
\newcommand\pcc{{\rm cm}^{-3}}

\newcommand\thh{\tau_{\rm H2}}

\newcommand\beq{\begin{equation}}
\newcommand\eeq{\end{equation}}

\newcommand{\Myr}{{\rm Myr}}

\def\apss{Ap\& SS}
\def\apj{ApJ}
\def\apjl{ApJL}
\def\aap{A\& A}

\def\mnras{MNRAS}

\def\aj{{AJ}}
\def\apjs{ApJS}

\def\mnras{{MNRAS}}

\title[Measuring molecular cloud ages]{Molecular Cloud Evolution VI. Measuring cloud ages}
 \author[ V\'azquez-Semadeni et al.] {Enrique
 V\'azquez-Semadeni,$^1$\thanks{E-mail: e.vazquez@irya.unam.mx} Manuel
 Zamora-Avil\'es,$^{2,1}$ Roberto Galv\'an-Madrid,$^1$ \newauthor and
 Jan Forbrich$^{3,4}$
\\ 
$^1$Instituto de Radioastronom\'ia y Astrof\'isica, UNAM. Apdo. Postal 72-3 (Xangari), Morelia, Michoc\'an
 58089, M\'exico\\ 
$^2$Department of Astronomy, University of Michigan, 311 West Hall, 1085 S. University Ann Arbor, MI 48109-1107\\
$^3$Centre for Astrophysics Research, School of Physics, Astronomy and Mathematics, University of Hertfordshire, College Lane,\\ Hatfield AL10 9AB, UK\\
$^4$Harvard-Smithsonian Center for Astrophysics, 60 Garden Street, Cambridge MA 02138, USA}

\date{Accepted XXX. Received YYY; in original form ZZZ}

\pubyear{2017}

\begin{document}
\label{firstpage}
\pagerange{\pageref{firstpage}--\pageref{lastpage}}
\maketitle

\begin{abstract}

In previous contributions, we have presented an analytical model
describing the evolution of molecular clouds (MCs) undergoing
hierarchical gravitational contraction. The cloud's evolution is
characterized by an initial increase in its mass, density, and star
formation rate (SFR) and efficiency (SFE) as it contracts, followed by a
decrease of these quantities as newly formed massive stars begin to
disrupt the cloud. The main parameter of the model is the maximum mass
reached by the cloud during its evolution. Thus, specifying the
instantaneous mass and some other variable completely determines the
cloud's evolutionary stage. We apply the model to interpret the observed
scatter in SFEs of the cloud sample compiled by Lada et al.\ as an
evolutionary effect so that, although clouds such as California and
Orion A have similar masses, they are in very different evolutionary
stages, causing their very different observed SFRs and SFEs. The model
predicts that the California cloud will eventually reach a significantly
larger total mass than the Orion A cloud. Next, we apply the model to
derive estimated ages of the clouds since the time when approximately
25\% of their mass had become molecular. We find ages from $\sim 1.5$ to
27 Myr, with the most inactive clouds being the youngest. Further
predictions of the model are that clouds with very low SFEs should have
massive atomic envelopes constituting the majority of their
gravitational mass, and that low-mass clouds ($M \sim 10^3$-$10^4
\Msun$) end their lives with a mini-burst of star formation, reaching
SFRs $\sim 300$-$500\, \Msun$ Myr$^{-1}$. By this time, they have
contracted to become compact ($\sim 1$ pc) massive star-forming clumps,
in general embedded within larger GMCs.

\end{abstract}




\section{Introduction} \label{sec:intro}

The lifetime of molecular clouds (MCs) remains an active 
research topic in the study of the interstellar medium and star
formation, and most recent studies, both observational and
theoretical, place this lifetime at a few times $10^7$ yr for clouds
in the $10^5$--$10^6 \Msun$ mass range
\citep[e.g.,] [] {BS80, Kawamura+09, ZA+12, ZV14, Lee+16}. In addition,
several observational studies have suggested that the star formation
rate (SFR) of the clouds appears to increase over their lifetimes. For
example,  studies of young clusters embedded in moderate-mass
MCs ($\sim 10^4 \Msun$) \citep[e.g., ] [] {PS99, PS00, DaRio+10} have
shown that their age histograms contain a large majority of young
(1--2 Myr) objects, but also a tail of older (up to several Myr) ones
 suggesting an accelerating star-formation activity,  sometimes
followed by a subsequent decline
\citep[see also] [] {Povich+16, Schneider+18}.  In addition, \citet{Kawamura+09}
reported a clear evolutionary process over the lifetime of giant
molecular clouds (GMCs, of masses $\sim 10^5$--$10^6 \Msun$) in the
Large Magellanic Cloud, evidenced by the increasing number of massive
stars across the sequence of GMC ``classes'' proposed by those
authors. Finally, on the basis of the large scatter in the observed
star formation efficiency in Milky Way GMCs, \citet{Lee+16} have
concluded that the SFR in those clouds must also be
time-variable.  Numerical simulations of MC formation and
evolution also exhibit time-varying, increasing SFRs during their
early stages \citep[e.g., ] [] {VS+07, Hartmann+12}. Also, in the
presence of stellar feedback, at late times the SFRs reach a maximum
and begin to decrease again \citep[e.g.,] [] {VS+10,
Colin+13}. \citet{VS+17} have recently shown that the
simulations of \citet{Colin+13} in fact produce stellar age
histograms highly resemblant of the observed ones \citep{PS99, PS00,
DaRio+10}, and reproduce observed radial age gradients in clusters
\citep{Getman+14} as well as bottom-heavy IMFs in scattered regions of
massive star formation \citep{Povich+16}.

However, most existing models for the SFR in MCs \citep[e.g.,] [] {KM05,
PN11, HC11, FK12} are based on the assumption that the clouds are near a state
of virial equilibrium between turbulence and self-gravity and are
therefore in a stationary state. They make predictions for an equally
stationary quantity, the star formation efficiency per free-fall time
($\eff$, the fraction of a MC's mass that gets converted into stars per
average free-fall time of the cloud) as a function of the parameters of
the turbulence. This type of models may be adequate for
predicting time- or space-averaged values of the SFR, but cannot
describe the evolution of the SFR in individual clouds if these evolve.

A different
class of models has been presented by \citet[] [hereafter Paper I]
{ZA+12}, \citet[] [hereafter Paper II] {ZV14}, \citet{Lee+16},
\citet{Volschow+17}, and \citet{Burkhart18}, who
have specifically included the time dependence of
the SFR. In particular, in Papers I and II we presented a model of
molecular cloud evolution (hereafter, the ZV14 model), in which we
assumed that MCs are in general formed by converging flows ({\it not}
collisions of pre-existing clouds) in the warm neutral medium. The
collisions produce layers of cold, dense atomic gas through nonlinear
triggering of the thermal instability \citep[e.g.,] [] {BP+99, HP99,
WF00, KI00, AH05, Heitsch+05, VS+06}. These layers start out thin, and
grow in thickness (and surface density) at constant volume density
\citep{VS+06} until they become Jeans unstable and begin to contract
gravitationally \citep{VS+07, Heitsch+08}. Thus, in the ZV14 model, we
assumed that the clouds begin to
undergo gravitational collapse  as soon as they reach their
thermal Jeans mass, having started from cold atomic gas
conditions.  However, the collapse is slow during the early
stages \citep[e.g., ] [] {BH13} and moreover clouds continue to accrete
mass from the converging flows. Thus, the clouds generally reach masses
significantly larger than their thermal Jeans mass.  

The immediate implication of the assumption of cloud contraction in the
ZV14 model is that the SFR of the clouds must be increasing over
time.  Theoretically, this can be understood in the sense that,
as the cloud contracts, its mean density increases, and therefore the
fraction of mass at high densities (i.e., short free-fall times) also
increases. This high-density tail of the density distribution is the
source of the ``instantaneous'' SFR of the cloud in the ZV14 and other
\citep{KM05, PN11, HC11, FK12} models. Thus, in the ZV14 model, as the mean
density increases, so does the mass fraction undergoing instantaneous
collapse, and the SFR increases. Paper I showed that the predicted
increase in the SFR was consistent with the observed age histograms in
embedded clusters \citep{PS99, PS00} and with the evolutionary sequence
for GMCs in the LMC proposed by \citet{Kawamura+09}, in both timescales
and stellar content. In Paper II it was furthermore shown that suitable
temporal averages of the ZV14 model reproduce the observed star
formation rates of nearby MCs, while ensemble averages, with an
appropriate weighting by a cloud mass spectrum, reproduce the locations
of full galaxies in an SFR {\it vs.}  dense gas mass diagram
\citep{GS04, Lada+12}.

The ability of the time-dependent ZV14 model to predict the evolution of
several cloud properties simultaneously (see Paper II) suggests the
possibility of applying it to estimate the ages of MCs. This is possible
because the model predicts a one-parameter family of model clouds, where
the main parameter is the total system
mass\footnote{The model of course also depends on the turbulent
parameters but, because it assumes that the initial conditions are those
of the cold atomic gas, these properties are assumed to be fixed, and so
the only free parameter is the total mass of gas accreted by the cloud
from the warm diffuse medium.}; that is, the total mass in the
converging streams that eventually undergoes a transition to the cold
phase. The evolution of all the other relevant physical quantities of
the model clouds, such as instantaneous dense mass, density, size, SFR,
and star formation efficiency (SFE) are self-consistently solved by the
model, and so, if any two of those can be measured simultaneously, they
can constrain the model to determine its total mass and evolutionary
stage.

In the present letter we present such an application to the cloud
sample compiled by \citet[] [hereafter, LLA10] {Lada+10}. Those authors
presented infrared extinction data for a set of 11 nearby, relatively
low-mass MCs, which included an estimate of the total cloud mass (mass
within the $A_K = 0.1$ extinction contour), the fraction of ``dense''
gas mass (mass above the $A_K = 0.8$ contour, which they estimate
corresponds to the mass at densities larger than $10^4 \pcc$), and an
estimate of the instantaneous SFE of each cloud given by the ratio of
the number of young stellar objects to the total cloud
mass.\footnote{It is important to note that \citet{Evans+09}
also reported SFRs and SFEs for a cloud sample that significantly
overlaps with that of LLA10, but the SFRs reported by
\citet{Evans+09} are significantly larger than those of LLA10. This is
in part due to different extinction cutoff definitions and in part to
different conversions from $A_V$ to mass. Since our model considers the
mass of all of the cold gas without regard to whether it is atomic or
molecular, the lower extinction cutoffs of LLA10 are more representative
of the system described by our model.}

The data from LLA10 shows a very large scatter of observed SFEs, with a
factor of $\sim 50$ between the largest and the smallest reported
SFEs. Those authors note, however, that, similarly to what happens for
whole galaxies \citep{GS04}, the observed SFEs appear to be proportional
to the {\it dense} ($n\ga 10^4 \pcc$) gas mass fraction. Here, we show
that this scatter can be understood in terms of the clouds being in
different evolutionary stages, and provide estimates for their ages,
profiting from the fact that the compilation by LLA10 contains all the
necessary information to constrain the ZV14 model to predict the
instantaneous SFE and thus infer the clouds' age.

The paper is organized as follows: In Sec. \ref{sec:model} we give
a brief review of the ZV14 model and of the LLA10 data. In Sec.\
\ref{sec:results} we present a comparison of the observed SFEs of the
clouds with the values predicted by the model at the observed
combinations of total cloud masses and the dense gas masses, showing a
good match to within factors of a few. In Sec.\ \ref{sec:disc} we
discuss some limitations and implications of our model, as well as how
it compares to other models for the SFR in clouds. Finally, in Sec.\
\ref{sec:concls} we present a summary and draw some conclusions.

\section{The model}\label{sec:model}

In this section we provide a brief qualitative description of the ZV14
model. This model aims at representing the main mechanism of
GMC formation, namely the compression of diffuse warm gas from the
interarm region as it enters the gravitational potential well of a
stellar spiral arm under solar neighborhood conditions \citep [see,
e.g., the review by] [] {Molinari+14}. We refer the reader to Papers
I and II for a detailed discussion of the model equations.

The ZV14 model essentially tracks the evolution of the mass budget in
clouds that are born as the result of a nonlinearly triggered phase
transition from the warm to the cold neutral atomic medium (the WNM and
CNM, respectively) by transonic compressions in the WNM, as
routinely observed in numerical simulations of dense cloud formation
\citep[e.g.,] [] {PVP95, BP+99, HP99, KI02, Heitsch+05, AH05, VS+06}.
In such simulations, the convergence of the flows {\it nonlinearly}
triggers a phase transition from the WNM to the CNM. In this type of
flows, the size of the forming cloud is not given by the most unstable
scale of the thermal instability as in the linear case, but rather, by
the transverse scale of the compressive motion acting on the WNM,
because it coherently induces the transition over a large area,
producing a thin sheet of cold atomic gas \citep{VS+06}. Subsequently,
the cold gas sheet often fragments into smaller clumps, but the ensemble
of small clumps begins to contract gravitationally as soon as it gathers
a mass larger than its thermal Jeans mass \citep{VS+07}.

The large-scale compressions can be driven by either large-scale
gravitational effects (e.g., the stellar spiral potential or Parker
instabilities) or by generic large-scale turbulent motions in
the WNM. The mass flux into the cloud is assumed to
last for 25 Myr, and to be given by $\dot M = \pi \rho_{\rm W}
\sigma_{\rm W} R_{\rm cl}^2$, where $\rho_{\rm W} = 1\,
\pcc$ is the density of the WNM, and corresponds to the
mean density of the ISM in the solar neighborhood, which is at the
lower end of the thermally-unstable range of the atomic ISM
\citep[e.g.,] [] {Field+69, Wolfire+03}; $\sigma_{\rm W} =
10 \kms$ is the 
velocity dispersion also in the WNM, and $R_{\rm cl}$ is the radius of
the initial flattened, circular cloud. This radius, and the total
duration of the mass flow, determine the total mass accreted by the
cloud over its lifetime, $\Mtot$. However, this quantity is quite
elusive, and in fact may never be observed as the cloud's mass, since
cloud erosion by feedback may start before all the diffuse gas is
converted to dense gas.  Therefore, throughout the paper we
characterize the model clouds instead by the maximum mass they reach
during their evolution, $\Mmax$, which we use as the single control
parameter of the model hereinafter.

The cloud is assumed to have moderately supersonic turbulence
(sonic Mach number $\Ms \sim 3$, as suggested by the studies of
\citet{KI02}, \citet{Heitsch+05}, \citet{AH05}, and \citet{Banerjee+09},
and to start their existence as thin, mostly atomic clouds, similar to
those observed by \citet{HT03}. The clouds' mass grows by continuing
accretion of warm atomic material at roughly constant volume density
(but increasing their thickness and column density), as described in
\citet{VS+06}, until they become Jeans-unstable, and begin to collapse
\citep{VS+07, VS+09}. The collapse is followed numerically, assuming the
cloud has a flattened geometry and a constant thickness, similarly to
what is observed in numerical simulations \citep[e.g.,] [] {VS+07, VS+11,
HH08, Banerjee+09, Clark+12}.

The model explicitly assumes that the strongly supersonic motions
observed in MCs ($\Ms \ga 10$) are dominated by infall, so that
the motions corresponding to true turbulence remain at
a roughly constant level, dictated by both the accretion
\citep[e.g.,] [] {Hunter+86, Vishniac94, WF00, KI02, Heitsch+06,
VS+06} 
and the collapse flows \citep[e.g., ] [] {VS+98, KH10, RG12,
MC15}, assuming that the turbulent kinetic energy injected by the
accretion or the collapse counteracts turbulent dissipation to
maintain a roughly constant, moderate turbulence level. 
This is consistent with the observation in numerical simulations that the
turbulence generated by the accretion or the collapse is never
sufficient to halt or significantly delay the collapse 
\citep[e.g.,] [] {VS+07, HH08, Banerjee+09, IM+16, Murray+17}.

Thus, we assume a lognormal probability density function (PDF) for the
density field \citep{VS94} having a constant width, corresponding to the
assumed Mach number of the initial conditions ($\Ms = 3$). We
adopt the prescription by \citet{Federrath+08}, with a
compressible-to-solenoidal $b$ parameter corresponding to half the
energy in each type of modes. The collapse of the cloud, and the
corresponding increase in the cloud's mean density, are modeled by
continuously shifting the mean density implied by the PDF to the
instantaneous mean density of the cloud.


Similarly to what is done in other models for the SFR \citep{KM05, HC11,
PN11, FK12}, the instantaneous SFR is then computed by calculating the
mass at densities $n > \nsf$ and dividing it by the free-fall time at
that density. The density $\nsf$ is a free parameter of the model, which
was calibrated in Paper I by matching the SFR predicted by the model to
that observed in a numerical simulation of a cloud of similar mass. The
resulting value of $\nsf = 10^6 \pcc$ has remained fixed in all
subsequent applications of the model.

Once the instantaneous SFR is computed, the model is advanced in time to
compute the corresponding increment in the stellar mass, which is
subtracted from the dense gas mass. Given the total stellar mass at
this time, a standard IMF \citep{Kroupa01} is used to compute the
instantaneous number of massive ($M > 8 \Msun$) stars in the cloud, and then
the instantaneous mass ionization rate on the cloud is computed using the
prescription from \citet{Franco+94}. Over the corresponding timestep,
the ionized mass is then also subtracted from the cloud's mass, and the
cycle is repeated.

In summary, the model tracks the cloud's mass budget over
time, according to the symbolic equation (Paper I)
\beq
M_{\rm cl}(t)=\int_0^t \dot{M}_{\rm inf}(t')~ {\rm d}t' -
M_{*}(t)-M_{\rm I}(t),
\eeq
where $M_{\rm cl}(t)$ is the instantaneous cloud (i.e., dense, cold)
mass, $\dot{M}_{\rm inf}(t)$ is the mass accretion rate onto the cloud
from the WNM inflows,\footnote{Strictly speaking, the
accretion does not need to be due to warm diffuse gas, and may refer
to {\it any} kind of accretion. However, if the accretion consisted
mainly of dense, cold material similar to that of the cloud, then it
should also be part of the collapsing cloud, and a more natural way to
represent this in the model would be to consider it as part of the
dense gas mass, rather than part of the accretion.} $M_{*}(t)$ is the
instantaneous mass in stars, and $M_{\rm I}(t)$ is the total mass that
has been ionized by stellar feedback. The detailed expressions and
procedures to derive each one of the terms in this equation are given
in Paper I.

The model thus follows, as a function of time, and for a given total
mass reservoir, the evolution of the instantaneous dense mass, density
PDF, radius, mean volume density, SFR, and SFE, computed as
SFE$=M_*/(\Mcl + M_*)$. Note that, in general, $\Mcl \le
\Mmax$, since the instantaneous mass of the cloud starts from zero and
grows as it accretes material from the diffuse medium, and then begins
to decrease as the stellar feedback within it begins to erode it.

\begin{figure*}
\includegraphics[width=0.45\linewidth]{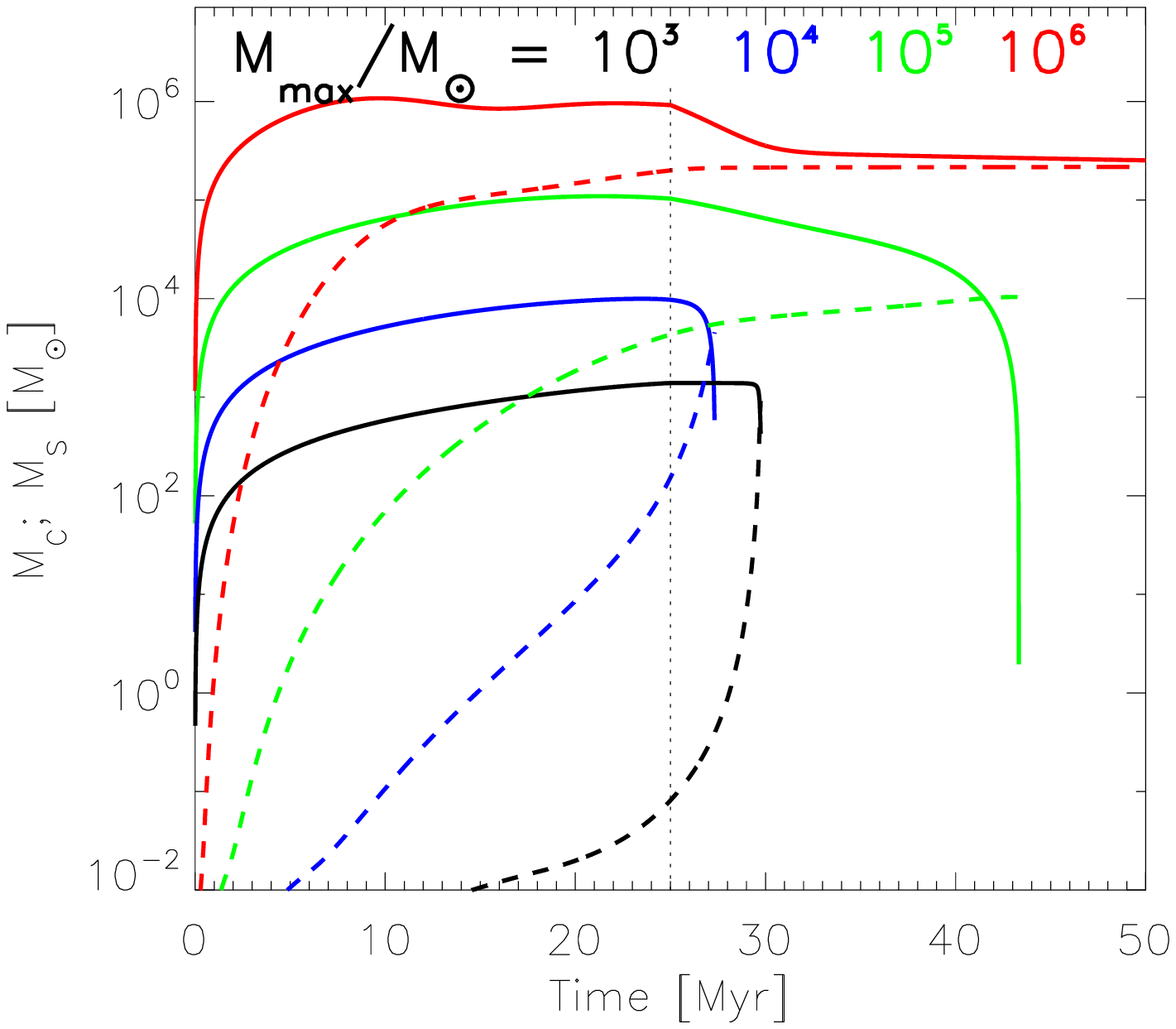}
\includegraphics[width=0.45\linewidth]{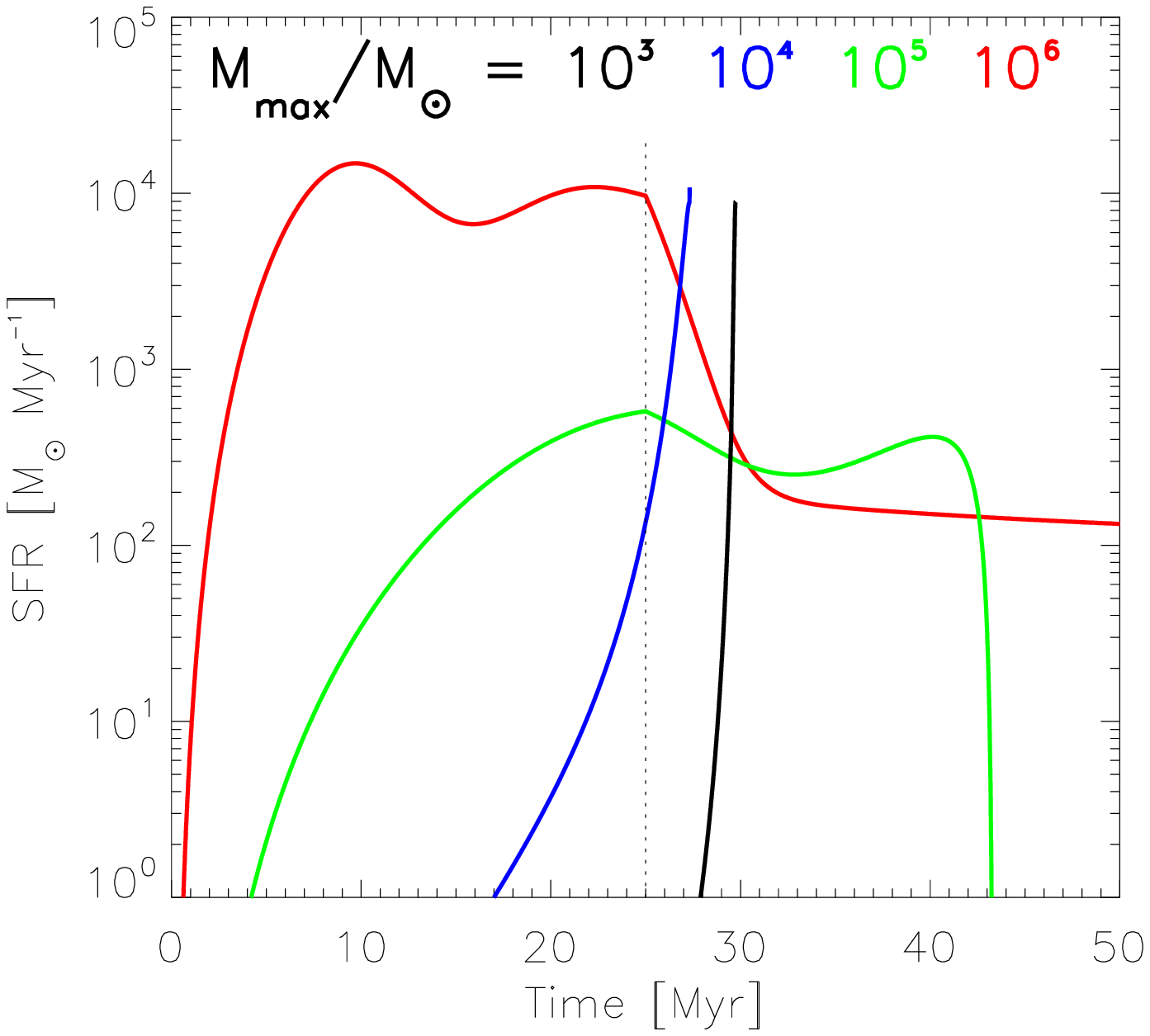}
\includegraphics[width=0.45\linewidth]{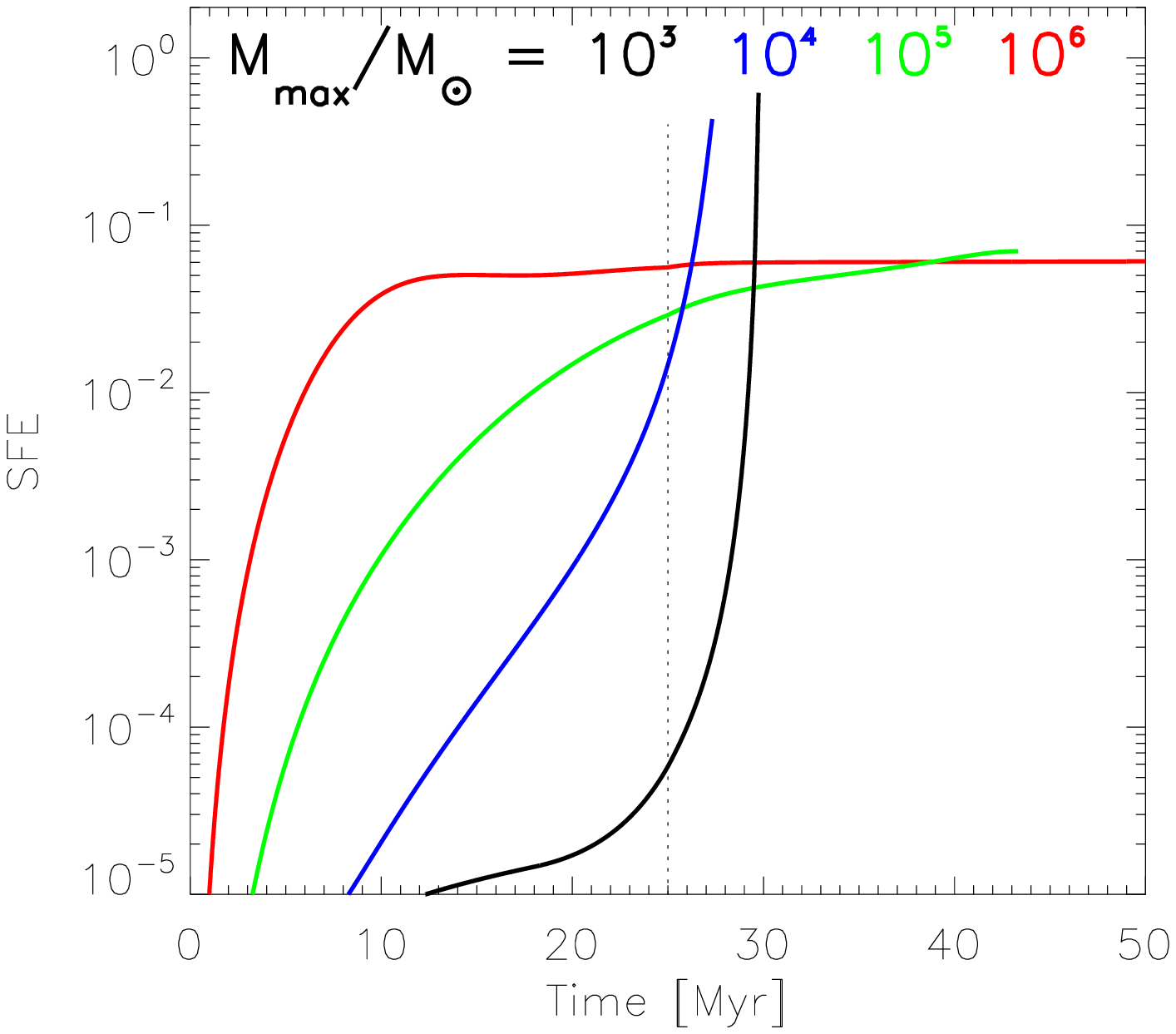}
\includegraphics[width=0.45\linewidth]{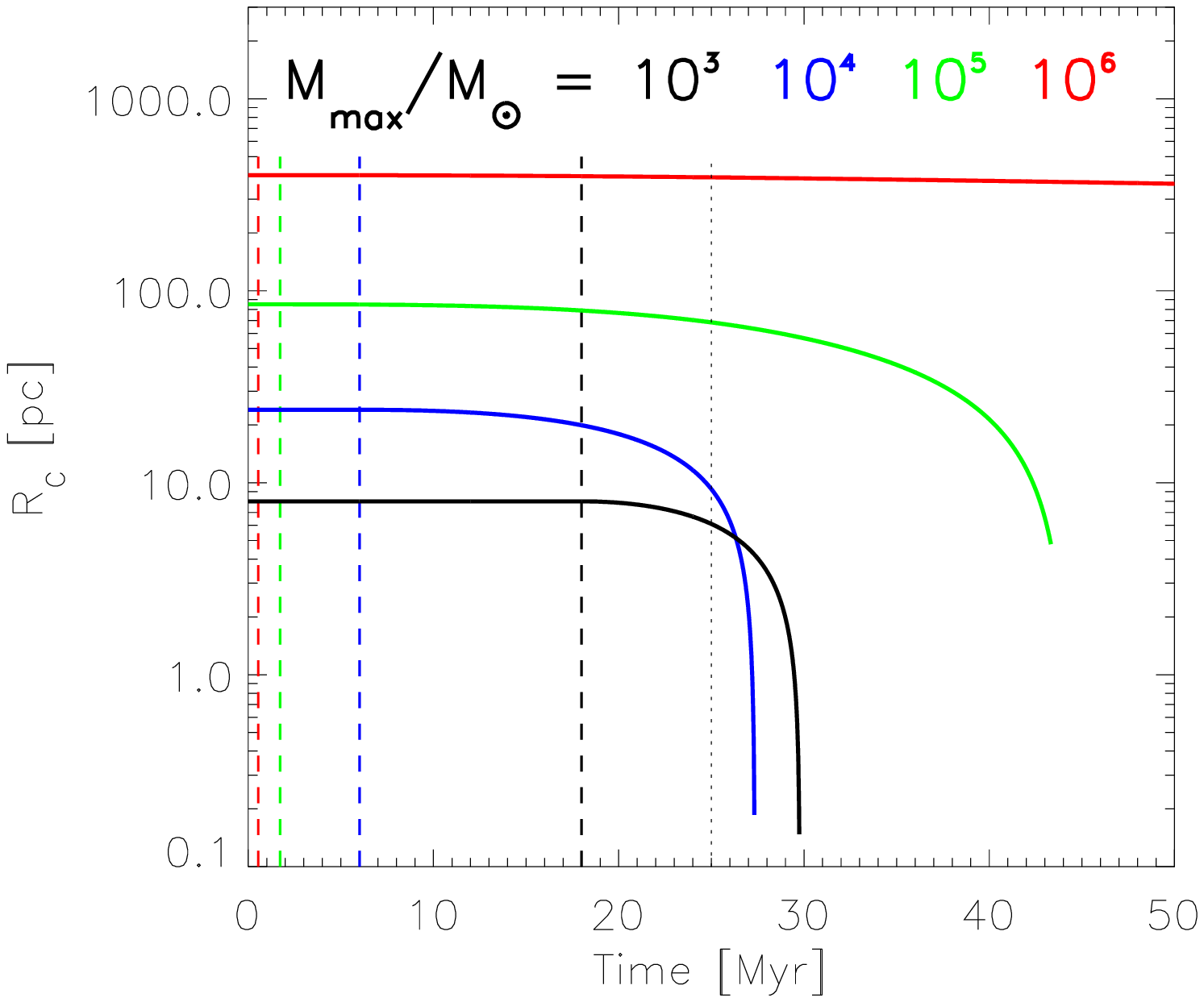}
\caption{Time evolution of the cloud mass and mass in stars (top left panel,
solid and dashed lines, respectively), SFR (top right panel), SFE
(bottom left panel) and radius (bottom right panel) for clouds with
$\Mmax=10^3$, $10^4$, $10^5$, and $10^6 \,
\Msun$ (black, blue, green and red lines, respectively). The
  vertical
dotted black line is the time at which the accretion stops ($t=25 \,
\Myr$). (Plots reproduced from Paper II).  } 
\label{fig:model}
\end{figure*}

In addition to these physical quantities, the
model can predict the instantaneous mass fraction of gas with density
larger than some threshold, under the assumption that the density PDF
retains its lognormal form, and simply shifts to higher densities as
the cloud contracts gravitationally.
For reference, we repeat in Fig. \ref{fig:model} the evolution
of the dense gas mass and the stellar mass ({\it left panel}), the SFR
({\it middle panel}) and the SFE ({\it right panel}) for clouds of
masses $10^3$, $10^4$, $10^5$, and $10^6 \Msun$ (black, blue, green, and
red lines, respectively), as first shown in Paper II. Note that, in
these figures, $t=0$ denotes the time when the WNM streams first
collide, and so the cloud has zero mass at $t=0$.

Note that the assumption that the density PDF remains lognormal is
  a questionable assumption of our model, since it is
now well known that the density PDF instead evolves by developing a
power-law tail at high densities \citep[e.g.,] [] {Kainulainen+09,
BP+11, Kritsuk+11, Girichidis+14, Lombardi+15, Lin+16}. However,
it has been suggested by
\citet{Kritsuk+11} that the power-law tail of the PDF at high densities is
the result of the development of highly peaked density profiles in
collapsing sites characterized by power-law radial
density profiles, which translate into a power-law density PDF. That
is, this tail is the result of the presence of {\it already collapsing
structures}. In our model, the density PDF represents the {\it turbulent
seeds} from which collapse starts, and so in Paper I we argued that the
relevant PDF is that of the turbulent fluctuations {\it before} they
begin to collapse, which is known to be lognormal \citep{VS94}, and is
the PDF routinely considered in models for the SFR and the IMF based on
the collapse of density fluctuations \citep{PN02, PN11, KM05, HC08,
HC11, Hopkins12, FK12}. Once the fluctuations begin to collapse, they
are already ``on route'' to forming stars, and thus they are already
counted by the model as a star that will form after a free-fall
time. Thus, their excess density should not be considered as the new
initial density of a subsequent collapse. For this reason, we opt for
assuming that the density PDF of the seed turbulent fluctuations retains
its original lognormal form, and representing the global collapse of the
cloud by a shift in the peak of the PDF to higher densities as dictated
by the increase in the mean density of the cloud.

Nevertheless, a recent study by \citet{Burkhart18} has
presented a model for the SFR similar to ours, but precisely taking
into account the development of a power law in the high-density range
of the PDF rather than shifting the entire PDF to higher densities as
we do, and obtaining similar results to ours. This suggests that, to
first order, the two methods for describing the evolution of the PDF
are roughly equivalent.

\begin{table*}
	\caption{Observed parameters of the cloud sample of \citet{Lada+10}} 
	\begin{tabular}{|c|c|c|c|c|c|c|c|c|}
		\hline
		Cloud  & Total mass$^1$ & Dense mass$^2$ & No.\ of YSOs & Observed SFE$^3$ \\
		name	& [$\Msun$]   & [$\Msun$]         &              & \%  \\
		\hline
		Orion A	&    67714      & 13721    & 2862 & 2.1  \\
		Orion B	&    71828      & 7261      & 635  & 0.44 \\
		California & 99930      & 3199      & 279  & 0.14 \\
		Perseus	&    18438      & 1880      & 598  & 1.6  \\
		Taurus	&    14964      & 1766      & 335  & 1.1  \\
		Ophiuchus &  14165      & 1296      & 316  & 1.1  \\
		RCrA    &     1137      & 258       & 100  & 4.2  \\
		Pipe    &     7937      & 178       & 21   & 0.13 \\
		Lupus 3 &     2157      & 163      & 69   & 1.6  \\
		Lupus 4 &     1379      & 124       & 12   & 0.43 \\
		Lupus 1 &     787       & 75        & 13   & 0.82 \\
		
		\hline
		$^1$Mass within the $A_K =0.1$ contour. \\
		$^2$Mass within the $A_K =0.8$ contour. \\
		$^3$According to eq.\ (\ref{eq:SFE_LLA10}). \\
		
	\end{tabular}
	\label{tab:LLA10_clouds}
\end{table*}

\section{Results} \label{sec:results}

As discussed in Sec.\ \ref{sec:model}, the ZV14 model predicts the
evolution of several physical quantities of a cloud of given total
mass as a function of time. Thus, in general, the model requires two
parameters (total mass and age) to be specified for a cloud in order
to completely determine its current evolutionary state. However, in
general, the age of the cloud is unknown, while several other
instantaneous quantities of the cloud, such as its instantaneous mass,
dense gas mass fraction, SFE, etc., are observables. Thus, any one of
those variables can be used as a proxy for time, in addition to its
instantaneous mass. That is, any combination of pairs of observables
(e.g., mass-dense mass fraction, mass-mean density, mass-radius, etc.) 
can constrain the instantaneous evolutionary state of a model cloud.

This capability of the model can then be used to test it against
observational data such as those by LLA10. These authors compiled data
on total cloud masses (i.e., mass above $A_K = 0.1$), dense gas masses
(i.e., mass above $A_K = 0.8$) and SFEs (given in that paper as the
instantaneous number of young stellar objects [YSOs] divided by the
clouds' mass) for a sample of 11 nearby clouds. We can thus use one
variable --- for instance, the dense gas fraction --- as the proxy for
age, use the mass to constrain the mass parameter of the model, and
then compare the observed SFE to that predicted by the model for a
cloud of the same instantaneous mass and dense mass fraction. If the
model passes this test, then the age it predicts for the cloud can be
taken as the actual physical age of the cloud. In what follows we
perform this procedure as a test for the model, and then apply it to
``date'' the clouds in the LLA10 sample.  For convenience, in the
second to fourth columns of Table \ref{tab:LLA10_clouds} we reproduce
the LLA10 data relevant for our study. In the fifth column we then
write the SFE implied by those data, defined as
\beq
{\rm SFE} = \frac{M_*} {M_* + \Mcl}
= \frac{0.5 N_{\rm YSO}} {0.5 N_{\rm YSO} + \Mcl},
\label{eq:SFE_LLA10}
\eeq
where $N_{\rm YSO}$ is the instantaneous number of YSOs, and $M_* = 0.5
\Msun N_{\rm YSO}$ is the total stellar mass, assuming that the mean stellar
mass is $0.5 \Msun$. The first equality is also the definition of the
SFE in the ZV14 model, since it computes the instantaneous stellar
mass during the evolution of a model cloud.

\begin{figure*}
\includegraphics[angle=-90, width=0.8\hsize]{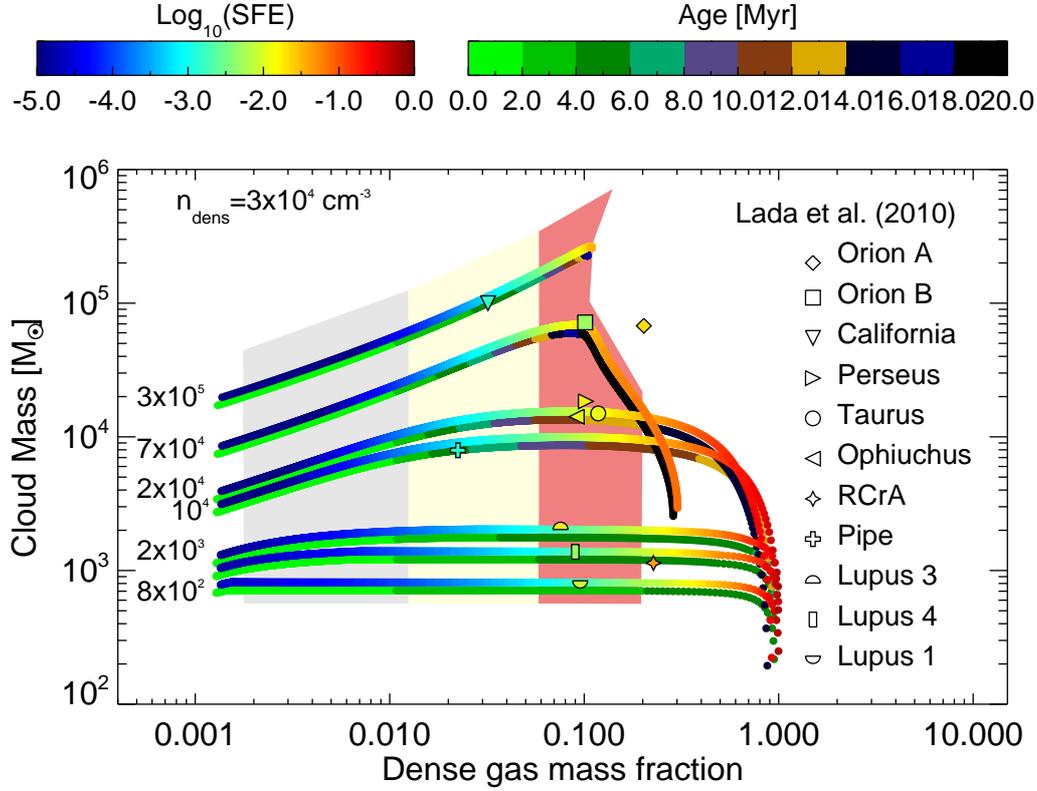}
\caption{Instantaneous cloud mass {\it vs.} instantaneous dense gas
mass fraction (i.e., mass at densities $n \ge 3 \times 10^4 \pcc$) of
model clouds of various total masses (indicated by the labels next to
each line). The tracks consist of two colored lines. The lines using the
color bar at the top left indicate the instantaneous SFE of the
model. The lines using the color bar at the top right show the cloud age
since becoming 25\% molecular, in 2-Myr intervals. The points show the
clouds from the LLA10 compilation, and their colors indicate the
reported SFE, using eq.\ (\ref{eq:SFE_LLA10}). The light blue, light
yellow and pink vertical bands respectively indicate the evolutionary
periods during which 25, 50, and 75\% of the gas mass is between $10^3
\pcc$ (left edge of each band) and $3\times 10^3 \pcc$ (right edge of
the band).
}
\label{fig:SFE_evol}
\end{figure*}

Figure \ref{fig:SFE_evol} shows the evolutionary tracks of model clouds
of various total masses (indicated by the labels next to each
line) in a diagram of instantaneous cloud mass {\it vs.} 
instantaneous dense gas mass (i.e., mass at densities $n \ge 3 \times 10^4
\pcc$), which we take as a proxy for LLA10's mass above $A_K = 0.8$,
and which we use as a proxy for the evolutionary time (or cloud
age). Each evolutionary track consists of two colored lines: one
showing the evolution of the SFE for each track, following the
continuous color bar shown at the top left of the figure, and one
giving the age of the model cloud since 25\% of its mass can be
considered molecular (see below), in 2-Myr intervals, following the
segmented color bar shown at the top right. Also shown in this plot
are symbols of various shapes corresponding to each of the clouds from
the LLA10 sample. Their location corresponds to the reported total and
dense gas masses, and their color corresponds to the reported SFE as
given in the 5th column of Table \ref{tab:LLA10_clouds}. This SFE can
thus be compared to the SFE of the nearest model evolutionary track at
the location of the point. We have chosen to show models whose
evolutionary tracks fall close to the location of the LLA10 data
points. 

It is clearly seen from Fig.\ \ref{fig:SFE_evol} that the color of the
points (the observed SFEs of the clouds in the sample) and those of the
model clouds at the locations of the points are similar, implying that
the observed SFEs of the clouds are consistent with their instantaneous
total and dense gas masses, as prescribed by our model.

\begin{figure}
\includegraphics[width=0.7\hsize, angle = 90]{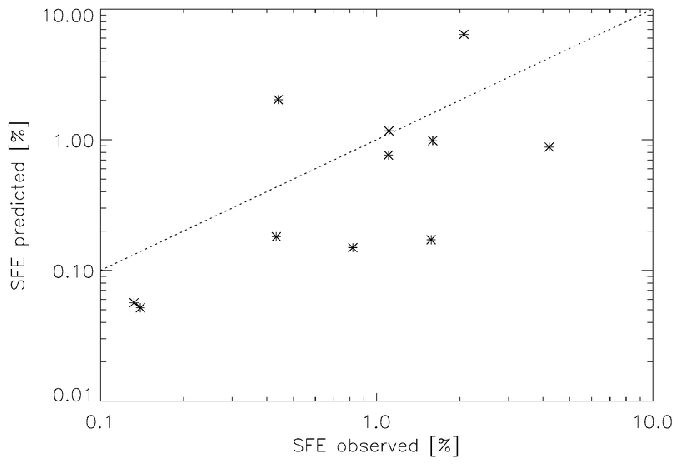}
\caption{Value of the SFE predicted by a model cloud that has the same
instantaneous cloud mass and dense gas fraction pair, $(\Mcl,\fden)$ as
each one of the clouds in the LLA10 compilation, plotted against the
corresponding observed value. }
\label{fig:sfe_obs_pred}
\end{figure}

To better quantify the degree of agreement between the predicted and
observed SFEs for the LLA10 clouds, in Fig.\ \ref{fig:sfe_obs_pred} we
plot the value of the SFE predicted by a model cloud that has the same
instantaneous cloud mass and dense gas fraction pair, $(\Mcl,\fden)$,
{\it versus} the observed value. The dotted line shows the identity
line. Although with significant scatter, a clear correlation is seen to
exist between the model-predicted and the observed values of the
SFE.  Thus, we propose that {\it the observed scatter in the
SFEs of the LLA10 clouds can be interpreted simply as a consequence that
the clouds are observed at different evolutionary stages.}

For reference, in Table \ref{tab:LLA10_clouds_modeled} we list 
a) the
maximum mass, $\Mmax$, of the model cloud with the same instantaneous
mass and instantaneous dense fraction;  b) the SFE of this model cloud at
the time when it has these values of the pair $(\Mcl,\fden)$; and
 c) the
predicted total age of this model; that is, the time since the moment
when the colliding streams first encountered each other.

\begin{table*}
\centering
\caption{Modeled parameters for the cloud sample of \citet{Lada+10}} 
\begin{tabular}{|c|c|c|c|c|c|}
\hline
Cloud      & $\Mmax$ model             & Predicted SFE & Time$^1$&  Molecular age$^2$ \\
           & [$\Msun$]                 & [\%]          & [Myr]      & [Myr]          \\
\hline
Orion A	   & $6.87 \times 10^4$ & 6.4            & 34.6         & 27.1 \\
Orion B	   & $6.87 \times 10^4$ & 2.0            & 24.3         & 16.8 \\
California & $2.65 \times 10^5$ & 0.052          & 5.47         & 2.2  \\
Perseus	   & $1.76 \times 10^4$ & 0.98           & 24.2         & 10.4 \\
Taurus	   & $1.55 \times 10^4$ & 1.2            & 24.6         & 10.1 \\
Ophiuchus  & $1.55 \times 10^4$ & 0.76           & 23.8         & 9.3  \\
RCrA       & $1.39 \times 10^3$ & 0.88           & 29.2         & 2.7  \\
Pipe       & $9.96 \times 10^3$ & 0.057          & 18.8         & 2.3  \\
Lupus 3    & $2.03 \times 10^3$ & 0.17		 & 27.6         & 2.5  \\
Lupus 4    & $1.39 \times 10^3$ & 0.18           & 28.6         & 2.1 \\
Lupus 1    & $8.16 \times 10^2$ & 0.15           & 33.4         & 1.6  \\

\hline
\end{tabular}
\begin{tablenotes}\footnotesize
	\small
	\item $^1$Total time since when the WNM streams first collide.
    \item $^2$Time since 25\% of the cloud's mass exceeded a density of
$3 \times 10^3 \pcc$.\\
\end{tablenotes}
\label{tab:LLA10_clouds_modeled}
\end{table*}

However, the time at which the WNM streams first encountered each
other is of little practical interest, since this event is unobservable.
A more interesting age is that since the cloud is already sufficiently
molecular to be identified as an MC. Although our model does not include
any chemistry, a first approximation to the molecular fraction can be
obtained by measuring the mass fraction above a density high enough that
the gas is most likely molecular there. We choose this ``molecular''
density as $\nmol = 3 \times 10^3 \pcc$. This value follows from the
standard prescription that the timescale for $H_2$ molecule formation is
$\thh \sim 10^9/n$ yr \citep{MM60}. Thus, at the density $\nmol$, the
timescale is $\thh \sim 3 \times 10^5$ yr, which is much shorter than
the timescales for MC evolution discussed here. Table
\ref{tab:LLA10_clouds_modeled} thus shows, in the fifth column, the
clouds' ages since they became significantly ``molecular''; i.e., since
25\% of their mass was at density $\nmol$ or larger. We see that the
clouds' ages according to this criterion range from $\sim 1.6$ to $\sim
27$ Myr. Most importantly, there is a general trend that the larger the
molecular age, the more efficiently the cloud is forming stars, as
prescribed by the ZV14 model.  It is worth noting that the ``molecular
ages'' of the lowest-mass clouds are in general lower than 3 Myr, which
may seem contradictory with the fact that some of those clouds (e.g.,
Lupus and Corona Australis) are known to have evolved class II young
stellar objects \citep[] [respectively] {Ansdell+16, CSA11}, which may
suggest an age larger than the one we measured. The correct way to
interpret this is that those objects formed when the mass of the clouds
was dominated by an atomic component, most likely in the form of an
atomic envelope, with still less than 25\% of the total cloud mass in
molecular form.

Finally, in Fig.\ \ref{fig:SFE_evol} we also show three colored vertical
bands that aim to provide a proxy for the evolution of the molecular
fraction ($\fmol$) of the gas, assuming that the gas becomes molecular
roughly when the volume density is between  1 and $3 \times 10^3
\pcc$.  Since the cloud is contracting gravitationally, it is
becoming denser on average, and thus the fraction of the cloud's mass
that is above a certain density threshold increases over time. The
light-blue band covers the interval between  the time when 25\% of
the gas mass is above $n = 10^3 \pcc$ (left edge of the bar) and when
25\% is above $n = \ 3 \times 10^3 \pcc$ (right edge). The light yellow
and pink bands show the corresponding evolutionary intervals for 50\%
and 75\% of the gas mass above these density thresholds.

\section{Discussion} \label{sec:disc}

\subsection{Implications and Insights} \label{subsec:implications}

Our results strongly suggest that the ZV14 model correctly describes, at
least to first order, the evolution of MCs and their SF activity, as a
consequence of their being in a state of global and hierarchical
collapse. The good average match between the observed SFEs and the
values predicted by our model for clouds of the same instantaneous mass
and dense mass fraction shows that the evolutionary state of a cloud can
be determined, at least to first order, when {\it a pair} of cloud
properties are known. This is because the evolutionary model constitutes
a one-parameter family of models, where the control parameter is the
total mass involved in the accretion and collapse process that forms the
cloud. Fortunately, although this total mass is in general unknown, it
suffices to know the {\it instantaneous} cloud mass, since, once it is
combined with another parameter, such as the dense fraction as we have
done here, it uniquely determines the both the total mass and the
evolutionary stage of the model. The model then allows to specify an age
for the cloud.

A number of points are worth noting. First, we remark that
the converging-flow setup is not essential for the evolution of the
SFR in our model. What determines this evolution is the process of
collapse. The converging flows are mostly important for {\it forming} a
cloud when there was none before. This is important to understand why
a cloud {\it begins} to collapse at some point. Otherwise, this point in
time (the onset of collapse) would be unconstrained. In the model,
this happens after the cloud reaches its thermal Jeans mass, due to
the accretion. But, since this accretion rate is assumed to remain
constant, it becomes progressively less important compared to the
increase in the cloud's mean density induced by the collapse as the
cloud evolves. This is especially true for the lower-mass clouds,
which have small cross sections for accreting diffuse gas. So,
the accretion becomes a secondary ingredient at late stages, and the
evolution proceeds towards increasing domination by the collapse.

Second, our result that the predicted and observed SFEs correlate well
for clouds of a given instantaneous mass and dense gas fraction could be
interpreted as to simply mean that the number of stars will be
proportional to the dense gas mass if the dynamical time for collapse in
that gas is constant.\footnote{We thank the referee for noting
this.} However, it is important to note that our model contains no
intrinsic assumption about any proportionality between the dense gas
mass and the number of stars, or the SFR. Rather, in the model, the
dense gas mass and the SFR are respectively given by the mass fraction
above the density threshold for defining "dense gas", $n_{\rm d} = 3
\times 10^4 \pcc$, and the mass fraction above the critical density for
star formation, $\nsf = 10^6 \pcc$. Both of these quantities depend on
the instantaneous values of the mean and standard deviation (assumed
constant) of the density PDF. The ratio between these two mass fractions
is not constant over time, because the density PDF is not linear with
density, and thus, as it shifts to higher densities over time, the ratio
of these two masses varies. Moreover, the instantaneous SFE is given by
the total stellar mass (which is proportional to the time integral of
the SFR) divided by the instantaneous cloud mass. The latter, in turn,
depends on the accretion, star formation, and mass loss rates. Thus, the
resulting match between the observed and predicted SFEs constitutes a
true test of the model, and not just the result of an imposed
proportionality between the dense gas mass and the number of stars.
Rather, this proportionality is then a prediction of the model.

Third, note that model clouds with different values of the
maximum mass may have the same value of their instantaneous cloud mass,
but at different evolutionary stages. For example, although the
instantaneous masses of the Orion A, Orion B and California clouds are
similar, the fact that the California cloud has a much lower SFE than
the Orion A cloud implies, according to the model, that the California
cloud system involves a larger total mass, but is at an earlier
evolutionary stage. Indeed, as seen in Table
\ref{tab:LLA10_clouds_modeled}, the model cloud that fits both Orion
clouds reaches a maximum cold-gas (cloud) mass of $\sim 6.9 \times
10^4 \Msun$, while the model cloud fitting the California cloud
reaches a maximum mass of $\sim 2.65 \times 10^5 \Msun$. This shows
that the evolution of MCs inherently relates the accretion onto the
cloud and its SF activity, since the mass growth of the clouds occurs
simultaneously with the increase in their SFR.

It is also important to remark that the model implies that the
transition of the cloud from being atomic-dominated to
molecule-dominated occurs gradually and simultaneously with the
increase in its star-forming activity. For example, it can be seen from
Fig.\ \ref{fig:SFE_evol} that low-SFE clouds such as the California and
the Pipe clouds are expected to be only roughly 50\% molecular (they lie
in the middle of the yellow band), implying that they should have about
50\% of their gravitational mass still in atomic form. This possibility
is usually overlooked when the gravitational binding of the clouds is
estimated via the clouds' molecular mass.

\subsection{The final mini-burst stages}
\label{sec:final_stg}

Finally, it is worth remarking that the model predicts quite large
SFRs and SFEs at the final stages of low-mass clouds. For example, it
is seen from the middle and right panels of Fig.\ \ref{fig:model} that
the model clouds with $\Mmax = 10^3 \Msun$ and $10^4 \Msun$
(respectively, the black and blue curves) reach peak SFRs $\sim 10^4
\Msun$ Myr$^{-1}$, and final SFEs $\sim 40\%$ and $60\%$,
respectively. These are in general not associated with low-mass
clouds. However, it should be kept in mind that the model follows the
evolution of the gas mass throughout its evolution, from the cold atomic
cloud stage \citep{VS+06} to the HII region stage. The large final
SFRs and SFEs correspond to stages when a few OB
stars and developed HII regions must be present. 

This can be exemplified by comparing the model prediction with a
strongly active region, such as the OMC-1 clump and its associated
Orion Nebula and the Orion Nebula Cluster (ONC). According to the data
collected in \citet{VS+09}, the OMC-1/ONC system  has a size
$\sim 1.2$-1.5 pc, contains a cold gas
mass $\Mgas \approx 2200
\Msun$ \citep{Bally+87} and $\sim 1600$ stars \citep{Tobin+09},
implying a stellar mass of $500$-$800 \Msun$, assuming a mean stellar
mass $M_{*} = 0.3$-$0.5 \Msun$. Moreover, its estimated age is $\la 2$ Myr
\citep{Hillenbrand97}. Therefore, this system has had an average SFR
of 250-400~$\Msun$ Myr$^{-1}$ over the last 2 Myr, and has a present
observed SFE $\sim 25$-$33\%$.  A recent estimate for the SFE of the OMC-1/ONC system by
\citet{DaRio+14} based on estimates of the free-fall time implied by the
mass distribution yields an SFE $\sim 30$-50\%.

This can be compared to the evolution of the $10^4$-$\Msun$ model cloud
shown in Fig.\ \ref{fig:model}. From the left panel of this figure, it
can be seen that at $t \approx 26.8$ Myr, $\Mstar \approx 10^3 \Msun
\sim 1/3 \Mgas$, for an SFE$=\Mstar/(\Mgas+\Mstar) \approx
25\%$.  Moreover, from the bottom-right panel of Fig.\
\ref{fig:model}, we see that the cloud's radius is decreasing very
rapidly, and has a size of a few parsecs.

To compute the average SFR of this model cloud, $\sfravg$, over the last
2 Myr, we note from the middle panel of Fig.\ \ref{fig:model} that, over
the time interval $24.8 \le t < 26.8$ Myr, the SFR may be approximated
by an exponential function of time. To estimate the characteristic
timescale of this function, we note that at $t \approx 26.8$, the star
formation rate of the $10^4$-$\Msun$ model is ${\rm SFR}(t=26.8) \sim
3\times 10^3\, \Msun$ Myr$^{-1}$, while 2 Myr earlier, ${\rm SFR}(t=24.8)
\sim 100\, \Msun$ Myr$^{-1}$. Fitting a straight line in log-lin SFR-$t$
space, we find
\[
{\rm SFR}(t) \approx {\rm SFR}(t_0) \exp\left(\frac{t-t_0} {\tau}
\right),
\]
where $\tau \approx 0.59$ Myr. Averaging this function over the time
interval $24.8 \le t < 26.8$ Myr, we obtain $\langle {\rm SFR} \rangle
\approx 880\, \Msun$ Myr$^{-1}$.

This estimate is 2-3 times larger than the observed $\langle {\rm SFR}
\rangle$ of the OMC1/ONC system. However, due to the steepness of the
SFR$(t)$ curve, this estimate is highly sensitive to the choice of time
interval. For example, if the starting point of the averaging interval
is taken as $t=24$ Myr instead of 24.8, the resulting $\langle {\rm SFR}
\rangle$ is $\sim 470\, \Msun$ Myr$^{-1}$, suggesting that, within the
uncertainties, the evolution of the SFR described by our model is
roughly consistent with observations. Moreover, we note that the SFE,
which is the result of the integrated SF activity over the evolution of
the cloud, is fully consistent with that observed for the OMC-1/ONC
system. Thus, we conclude that the final SF burst of the low-mass
regions predicted by our model adequately describes the evolution of
these systems.

\subsection{Assumptions and limitations} \label{sec:limitations}

Our model is of course subject to a number of assumptions that limit its
predictive ability to only order-of-magnitude precision. 
Besides the assumption of a persistent lognormal
PDF discussed in Sec.\ \ref{sec:model}, which may or may not be a
problem, another limitation of our model is that it only considers
collapse and cloud destruction by photoionising radiation. It neglects
possible delay of the collapse by magnetically supercritical magnetic
fields, additional cloud destruction/dispersal processes such as
supernovae, stellar winds, etc., and, particularly importantly,
variations in the accretion rate due to processes other than the
inertial mass flux we have considered. All of these mechanisms may be
responsible for the significant scatter observed in the plot of
predicted-{\it vs.}-observed SFE (Fig.\
\ref{fig:sfe_obs_pred}). Another source of uncertainty is that we have
used a volume density threshold ($3 \times 10^4 \pcc$) for comparison to
an column density one (LLA10's $A_K = 0.8$ definition of high column
density gas), and the correspondence between the two types of density is
far from perfect. Nevertheless, using high-volume density gas is
actually closer to the physical motivation behind the consideration of
high-column density gas, since LLA10 themselves assume that the $A_K >
0.8$ gas is representative of gas with $n > 10^4 \pcc$, on the basis of
the assumption that it is the dense gas that is actually responsible for
star formation.

Another limitation of our model, in its application for the
present study,\footnote{The initial physical conditions of the model
may be specified at will, and, in fact, applications to different
environments can be achived by specifying the appropriate initial
conditions for each environment, such as the Central Molecular Zone of
the Milky Way, for example.} is that we have assumed that all clouds
start from the same initial conditions, namely those of the CNM in the
solar neighborhood, and with the same accretion rate from the WNM.
Fluctuations in these initial conditions, in particular in the mean
density and temperature of the forming clouds, will cause fluctuations
in the clouds' thermal Jeans mass, and therefore in the time of the
onset of collapse. This effect surely contributes to the scatter we
observe in the predicted-{\it vs.}-observed SFE plot of Fig.\
\ref{fig:sfe_obs_pred}.

Finally, yet another idealization of our model is the assumption that
the accretion onto the clouds consists exclusively of warm diffuse
gas. This is a reasonable first-order approximation for solar
neighborhood conditions, as it is known that, at the solar
galactocentric radius, the azimuthally-averaged molecular mass
fraction is only 10-20\%, and the gas cycles from predominatly atomic
to molecular as it passes through the spiral arms \citep[e.g.,] []
{Koda+16}. Moreover, since the mean density of the atomic gas at the
solar radius is $n_{\rm H} \sim 1\, \pcc$ \citep[e.g.,] []
{Ferriere01}, this gas is predominantly in the warm phase. Thus, the
gas from which the GMCs in the solar neighborhood form is expected to
be WNM.

Nevertheless, in reality, even if the accretion onto the GMCs consists
of predominantly-diffuse gas from the interarm region as it enters a
spiral arm, it is likely to contain a ``mist'' of dense, cold cloudlets.
This is because the dense gas seems to not be fully destroyed by stellar
feedback as it exits the previous spiral arm. Instead, only part of it
is truly destroyed, while the rest is dispersed into smaller units
\citep{Koda+16}. Thus, a more realistic description of the assembly of
GMCs would include this mist of cold clumps.

The problem of GMC assembly by diffuse-gas streams containing
scattered cold clumps has been investigated numerically by
\citet{Carroll+14}.  These authors compared two converging-flow
simulations, in both of which the mean density of the inflows is
$\langle n \rangle = 1\, \pcc$, but being uniform in one case, and
clumpy in the other. In the latter, there is a substrate of density $n
= 0.25\, \pcc$ and a mist of clumps of radius 0.55 pc and density
$n_{\rm c} = 15.2\, \pcc$. They found that, in the clumpy run, the
forming cloud fragments less, collapses later, and acquires more mass,
because the substrate's density is lower, implying higher temperatures
in the compressed layer, and thus a larger Jeans mass. The clump-clump
collisions are not very efficient because their collisional
cross-section is small. Thus, once the cloud becomes gravitationally
unstable, the global collapse is more focused, and the SFR reaches
higher values than in the smooth run, leading to higher final total
stellar mass, by a factor of $\sim 2$. Thus, the presence of dense
cloudlets does not significantly affect the evolution of the cloud.

\section{Conclusions} \label{sec:concls}

In this paper we have shown that the our ZV14 evolutionary model of
collapsing clouds and their SFR captures to order-of-magnitude
precision the observed scatter in the SFE of MCs, and provides an
interpretation of it in terms of different clouds being at different
evolutionary stages, since the model predicts that the SFR of the
clouds varies in time, first increasing as the clouds' density
increases during collapse, and then decreases as stellar feedback
begins to disrupt the clouds. This interpretation is consistent with
previous works proposing that an ``uncertainty principle'' applies to
observations of the SFR in external galaxies below a certain spatial
scale, because the observed regions are small enough that local
evolutionary differences cannot averaged out over the region, and
different regions are caught in different evolutionary states, so that
they will display different SFRs at a given gas mass \citep{KL14,
Kruijssen+18}.

The fact that our model correctly captures the average evolutionary
trend of the SFE with other cloud parameters suggests that the
dominant mechanisms controlling MC evolution are indeed their global
gravitational collapse and their subsequent destruction by stellar
feedback, as described by our model, with other processes providing
second-order corrections. These are better followed by detailed
numerical simulations. However, the model allows an understanding of
the fundamental physical underlying processes.

Also, our results imply that the reason higher-density gas appears to
correlate linearly with the SFR, while lower-density gas exhibits a
looser correlation \citep[e.g.,] [] {GS04, Bigiel+08, Lada+12} is not
because only the dense gas forms stars, but because it is {\it closer}
in time and space to forming stars than the lower density gas, as
suggested by \citet{BH13}. In turn, this occurs because of the global
collapse of MCs we have proposed \citep{VS+09}, since gravitational collapse is
in general extremely non-homologous, amplifying density gradients, and
causing an accelerating gas flow from the low- to the high-density
regions. 

We stress that, in our model, the clouds do not have a
well-defined time at which they are ``born'', since their molecular
(i.e, dense gas) fraction increases over time. This in turn implies
that, especially during the early evolutionary stages of the clouds, the
dynamic role provided by the weight of its atomic envelope is
important, and cannot be neglected when considering the gravitational
boundedness of a cloud.

Finally, our model predicts that low-mass clouds ($M \sim
10^3$-$10^4\, \Msun$) undergo a strong mini-burst of SF at the end of
their lives, when they constitute a compact, massive {\it clump},
generally embbeded within a larger, more massive cloud. Although this
prediction may appear as counterintuitive at first, because low-mass
clouds are in general associated with low SFRs, it must be understood in
terms of an {\it evolutionary} sequence. Although at its initial stages
the cloud has sizes $\sim 10$ pc, densities of a few times $100\,
\pcc$, and SFRs $\sim 10\, \Msun$ Myr$^{-1}$, and therefore corresponds
to our standard definition of a ``low-mass cloud'', by the time such a
gas parcel reaches its final stages, it has contracted to sub-parsec
scales and reached densities $n \ga 5\times 10^3\, \pcc$ (see Fig.\ 1 of
Paper II), with SFRs $\ga 300\, \Msun$ Myr$^{-1}$, thus corresponding to
our notion of a ``massive clump''. Moreover, since accretion onto the
cloud is expected to continue from its environment, this clump is now
expected to be part of a larger-mass system, which would correspond to a
larger-mass cloud in our model. Thus, the model proposes a unification
of the various classes of objects into a general evolutionary picture in
which all of the cloud properties, such as its mass, density, size and
star formation activity change in time, transiting from quiescent to
bursting stages, and then being destroyed by the stellar
feedback. Further testing and predictions of the model will be presented
in future contributions.

\section*{acknowledgements}
We thankfully acknowledge an anonymous referee for providing an
insightful report that prompted us to clarify several aspects of our
model, and C.\ Lada and D.\ Kruijssen for useful comments on our
manuscript. This work has received partial financial support from
CONACYT grant 255295 to E.V.-S. R.G.-M.\ acknowledges support from
UNAM-PAPIIT program IA102817. J.F.\ wishes to acknowledge the Visiting
Professor Program from the Graduate School at UNAM.


%
%
%
\bsp	
\label{lastpage}
\end{document}